\documentclass[pra,twocolumn,eqsecnum]{revtex4}
\usepackage{amssymb, amsbsy, amsmath, latexsym, dsfont, array, layout, graphicx}
\usepackage{color}

\newcommand{\ket}[1]{\left|{#1}\right\rangle}
\newcommand{\bra}[1]{\left\langle{#1}\right|}

\begin{document}
\title{Quantum walk on circles in phase space}
\author{Peng Xue}
\author{Barry C. Sanders}
\affiliation{Institute for Quantum Information Science, University of Calgary, Alberta T2N 1N4, Canada}

\begin{abstract}
We propose a variation of the quantum walk on a circle in phase space
by conjoining the Hadamard coin flip with
simultaneous displacement of the walker's location in phase space and show that this generalization is a proper quantum walk albeit over multiple concentric circles in phase instead of just over one circle. We motivate the
conjoining of Hadamard and displacement operations by showing that the Jaynes-Cummings model for coin+walker
approximately yields this description in the dispersive limit.
The quantum walk signature is evident in the phase
distribution of the walker provided that appropriate pulse durations are applied for each coin flip.
\end{abstract}


\maketitle

\section{Introduction}
\label{sec:introduction}

The quantum walk (QW) is one of the most important developments in theoretical quantum information science, both as an
intriguing generalization of the ubiquitous random walk (RW) in physics~\cite{Aharonov,Kempe} and for exponential algorithmic
speed-ups~\cite{Chi03,bit,glued,NAND}. The quincunx, or Galton Board~\cite{Galton}, was developed to exhibit the features of
random walks in experiments and more recently an optical quincunx that simulates a `wave walk'~\cite{Bou99} was demonstrated.

For the quantum quincunx, an appealing strategy for experimental realization arises in the context of a QW over a circle in
phase space \cite{Tra02}, which arises naturally for a simple harmonic oscillator. Points in phase space correspond to the
oscillator position-momentum pair $(x,p)$, which we henceforth refer to as the phase space `location', and energy-conserving
evolution of the oscillator guarantees that $E=(x^2+p^2)/2$ (for the oscillator of unit mass and unit frequency) is a
conserved quantify, thereby constraining the phase space trajectory to circle in phase space centered at the origin $(0,0)$.

The discrete walk on the circle corresponds to phase jumps $\Delta\theta=\theta_2-\theta_1$
for
\begin{equation}
\label{eq:theta}
    \theta_i=\tan^{-1}p_i/x_i,
\end{equation}
which is well-defined provided that $x\neq 0 \neq p$. The discrete random walk on the circle, corresponding phase jumps
$\pm\Delta\theta$, with $\Delta\theta$ of fixed size and the sign $\pm$ chosen randomly, has been used to provide a clear
explanation of phase diffusion of the laser field~\cite{Swa99}. More recently the random walk on the circle in phase space has
been generalized to the QW on a circle in phase space: in the quantum case the walker's location as a point in phase space is
replaced by a localized wave function centered at a location $(x,p)$, and the random flip of sign $\pm$ is replaced by a
quantum coin given by a qubit, which is flipped by a Hadamard operation and then entangled with the oscillator by free
evolution. An example of a localized wave function is the coherent state
\begin{equation}
\label{eq:coherentstate}
    |\alpha\rangle=D(\alpha)|0\rangle
\end{equation}
with~$|0\rangle$ the ground state of the simple harmonic oscillator and
\begin{equation}
\label{eq:displacement}
    D(\alpha)=\exp(\alpha \hat{a}^\dagger-\alpha^*\hat{a})
\end{equation}
the unitary displacement operator~\cite{Glauber}, and $\alpha=(x+\mathrm{i}p)/\sqrt{2}$ for localization at~$(x,p)$. The full
quantum walk on a circle in phase space is described in detail in Sec.~\ref{sec:background}.

In these quantum-walk-on-the-circle schemes, the coin qubit is directly controlled; however, in the context of cavity quantum
electrodynamics, the coin qubit is an atom within a high-finesse resonator. The high-finesse nature of the cavity mitigates
against direct control of the atom, and the only viable option is to drive the coin qubit indirectly by driving the cavity,
which in turn drives the atom. Here we show that this indirect coin flip indeed suffices to create a quantum quincunx, but the
quantum walk is no longer confined to a circle in phase space but rather undergoes a quantum walk that hops between different
circles in phase space. In Sec.~\ref{sec:circles} we explain this revised QW involving simultaneously driving of both the
oscillator and the coin qubit.

In Sec.~\ref{sec:JC2QW} we consider the Jaynes-Cummings (JC) model~\cite{JC} as the underpinning of the results in
Sec.~\ref{sec:circles}. Whereas Sec.~\ref{sec:circles} presents a generalized Hadamard transformation of the coin that
involves driving both the oscillator and the coin, in this section we use the JC model, which is one of the most important
models in quantum optics to describe cavity quantum electrodynamic systems, and was originally used to explain the maser (and
hence the laser)~\cite{JC}. In the dispersive limit, and with judicious timing to achieve the right phase steps, we show in
this section that a QW on circles in phase space can be well approximated by JC dynamics. The results are summarized in
Sec.~\ref{sec:conclusions}.

\section{Background}
\label{sec:background}

The random walk on the circle in phase space, used to describe laser diffusion~\cite{Swa99}, comprises two coupled systems:
the walker, who is physically a simple harmonic oscillator, and the unbiased two-sided coin, which is mathematically an
unbiased random bit. The joint system of the coin+walker has a state space $\mathcal{L}^1(\mathbb{R})\times\{0,1\}$. That is,
the walker's state corresponds to distributions in $\mathcal{L}^1(\mathbb{R})$, and the coin can have either value
$\varsigma\in\{0,1\}$. Evolution consists of alternating coin flips, which generates $0$ or $1$ randomly with equal
probability, and then the walker's distribution in phase space is rotated by an angle $\pm\Delta\theta$ with the sign $\pm$
given by $(-1)^\varsigma$.

In quantizing the QW, the walker's distribution is replaced by a state $\rho\in\mathcal{B}(\mathcal{H}_\text{w})$ for
$\mathcal{B}(\mathcal{H}_\text{w})$ the Banach space of bounded operators on
$\mathcal{H}_\text{w}\cong\mathcal{L}^2(\mathbb{R})$. The coin is replaced by a qubit with Hilbert space
$\mathcal{H}_\text{c}\cong P\mathbb{C}^2$, namely the projective space of two-component complex vectors. The joint coin+walker
space $\mathcal{H}_\text{c}\otimes\mathcal{H}_\text{w}$ is spanned by a basis set comprising tensor products of Fock states
\begin{equation}
\label{eq:Fock}
    |n\rangle=\frac{\hat{a}^{\dagger n}}{\sqrt{n!}} |0\rangle,\,
    \hat{a}^\dagger=2^{-1/2}\left(\hat{x}-\mathrm{i}\hat{p}\right)
\end{equation}
for $|0\rangle$ the oscillator ground state in Eq.~(\ref{eq:Fock}),
and $|0\rangle$ and $|1\rangle$ the two coin basis states.
The Fock states are also known as number states, and Fock state $|n\rangle$ is
an eigenstate of the number operator $\hat{n}\equiv\hat{a}^\dagger\hat{a}$ with
eigenvalue~$n\in\mathbb{N}$.

The QW is effected as an alternating sequence of two operations, namely the Hadamard transformation on the coin
\begin{equation}
\label{eq:H}
    H=|+\rangle\langle 0| + |-\rangle\langle 1|,\,
    |\pm\rangle=\left(|0\rangle\pm |1\rangle\right) /\sqrt{2},
\end{equation}
and the free evolution
\begin{equation}
\label{eq:F}
    F(\Delta\theta)=\exp\left(\mathrm{i}\hat{n}\hat{\sigma}_{z}\Delta\theta\right)
\end{equation}
between coin flips. The free evolution effects a conditional rotation of the walker's state by an angle $\pm\Delta\theta$
which is chosen given an initial walker state $\ket{\alpha}$ \cite{San03}:
\begin{equation}
\frac{1}{\sqrt{\bar{n}}}<\Delta\theta<\frac{2\pi}{\bar{n}+\sqrt{\bar{n}}} \label{eq:deltatheta}.
\end{equation} In fact the
evolution of the walker can be entangled with the coin state by this evolution, and this entanglement between the coin and
walker degrees of freedom underpins the dramatic differences between the classical random walk vs the QW. The resultant
evolution is achieved by repeated application of the QW unitary operator
\begin{equation}
\label{eq:U}
    U=F(H\otimes\openone);
\end{equation}
after $N$ discrete time steps,
the state of the coin+walker evolves according to the evolution operator $U^N$.


As we shall see, the QW signature will be evident in the phase distribution of the walker's state~\cite{San03,Di04}. The phase
distribution for the walker's reduced state $\rho_\text{w}$, obtained by tracing out the joint coin+walker state over the
coin's degree of freedom, is
\begin{equation}
\label{eq:phasedistribution}
    P(\phi)=\lim_{M\rightarrow\infty}\!_M\langle\phi|\rho_\text{w}|\phi\rangle_M
\end{equation}
as constructed from phase states~\cite{Lou73}
\begin{equation}
\label{eq:phasestate}
        \ket{\phi}_M\equiv\frac{1}{\sqrt{M}}
        \sum_{n=0}^{M-1}\mathrm{e}^{\mathrm{i}n\phi}\ket{n}.
\end{equation}
Phase states are thus dual to the Fock states in the sense that
\begin{equation}
    \langle n|\phi\rangle_M=\mathrm{e}^{\mathrm{i}n\phi}/\sqrt{M}
\end{equation}
if $n<M$,
and the overlap is zero otherwise, and,
for $\phi_m\equiv 2m\pi/M$,
\begin{align}
    \text{span}&\{\ket{\phi_m};m=0,1,\ldots,M-1\}\nonumber\\=&\text{span}\{\ket{n};n=0,1,\ldots,M-1\}
\end{align}
with $\{\ket{\phi_m}\}$ an orthonormal basis of the subspace. For arbitrary phase states $\ket{\phi}$ and $\ket{\phi+\delta}$,
their overlap is given by
\begin{align}
    _M\!\langle\phi|\phi+\delta\rangle_M
        &=\frac{1}{M}\sum_{m=0}^{M-1} \mathrm{e}^{\mathrm{i}m\delta}\\
        &=\frac{1}{M}\mathrm{e}^{\mathrm{i}(M-1)\delta}U_{M-1}(\delta)\nonumber
\end{align}
for
\begin{equation}
    U_{M-1}(\delta)=\frac{\sin(M\delta/2)}{\sin(\delta/2)}
\end{equation}
the Chebyshev polynomial of the second kind~\cite{Gra80}.

In the coin+walker basis, with phase states as the walker basis states,
the free evolution operator acts according to
\begin{equation}
    F\ket{\varsigma,\phi}=\ket{\varsigma,\phi+(-1)^\varsigma\Delta\theta}, \;
        \varsigma\in\{0,1\}
\end{equation}
so the phase states form a natural representation for studying this evolution. Furthermore the signature of both the random
walk and the QW, and their differences, is in the phase distribution~(\ref{eq:phasedistribution}) of the reduced walker
state~$\rho_\text{w}$.

The dispersion of the phase distribution is especially important. As moments are not particularly useful for distributions
over compact domains, other strategies are needed. For the phase distribution over the domain $[0,2\pi)$, Holevo's version of
standard deviation~\cite{Holevo} is particularly useful as it reduces to the ordinary standard deviation for small spreads and
is sensible when the dispersion is large over the domain~\cite{Wis97}. Holevo's standard deviation is
\begin{equation}
\label{eq:Holevo}
    \sigma_\text{H}=\sqrt{|\langle \mathrm{e}^{\mathrm{i} \phi}\rangle|^{-2}-1},\,
        \langle \mathrm{e}^{\mathrm{i} \phi}\rangle
            =\int_0 ^{2\pi} \mathrm{d}\phi P(\phi) \mathrm{e}^{\mathrm{i}\phi}
\end{equation}
with respect to any phase distribution $P(\phi)$ (\ref{eq:phasedistribution}).

The Holevo standard deviation has been shown to evolve according to $\sigma_\text{H}\propto t$ for the QW, whereas
$\sigma_\text{H}\propto \sqrt{t}$ for the RW, at least for short times where the phase distribution has support over less than
the circle~\cite{San03}. This quadratic speed-up of phase spreading in a unitary evolution is a hallmark of the QW on the
circle. We will use this quadratic speed-up as the indication of QW in the system.

Our focus is on phase spreading as a signature for the QW, but phase is not directly measurable. However phase can be inferred from homodyne or from optical homodyne tomography measurements~\cite{San03}.

\section{Quantum walks on circles}
\label{sec:circles}

In previous schemes, implementations of the QW on a circle have been proposed
for ion traps~\cite{Tra02} or cavity QED~\cite{San03,Di04}, and each scheme relies on
direct driving of the coin (i.e.~directly flipping the coin without modifying the cavity field).
In realistic systems this may not be possible, and instead the simple harmonic oscillator
will be driven, which then drives the coin via the oscillator-coin coupling.

\subsection{Generalized Hadamard Transformation}

In this section we treat this strategy of indirectly driving the coin
by generalizing the Hadamard transformation to
\begin{equation}
\label{eq:HD}
    H\mapsto \exp\left\{\mathrm{i}\frac{\pi}{4}\left[\hat{\sigma}_x
        +\lambda\hat{x}\right]\right\}
    =H\otimes D,
\end{equation}
for $D(\alpha)$ the unitary displacement operator~(\ref{eq:displacement}) with $\alpha\mapsto\mathrm{i}\lambda/\sqrt{2}$. Thus
$\lambda$ is the kick the walker receives during the Hadamard pulse. In the next section we will derive an approximation to
the unitary operator~(\ref{eq:HD}) by beginning with the JC model Hamiltonian.

\begin{figure}
\includegraphics[width=8.5 cm]{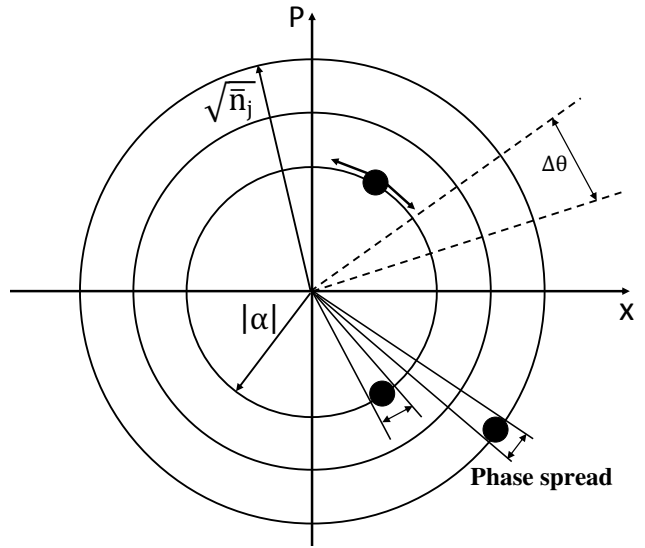}
   \caption{
   Phase space diagram, with coordinates $x$ and~$p$,
   depicting circles of fixed radius~$\sqrt{\bar{n}_j}$ for three different
   values of~$n_j$ corresponding to three circles indexed by~$j$.
   The large black dot on the innermost circle represents the distribution of the walker.
   The arrows extending from the large black dot represent a fixed phase jump $\pm\Delta\theta$,
   which is clockwise or counterclockwise depending on the sign of~$\Delta\theta$.
   The dashed line shows a jump size of~$\Delta\theta$ and the corresponding arcs of the
   circles subtended by this angle.
   The geometric meaning of the displacement operator parameter~$\alpha$
   is shown as the modulus being the radius of the $j^\text{th}$ circle; not shown is that
   arg$\alpha$ is the phase of the center of the large black dot.
   }
   \label{fig:blobby}
\end{figure}

The generalized Hadamard transformation~(\ref{eq:HD}) nicely factorizes into a Hadamard transformation and a displacement
operation. The Hadamard tranformation effects the desired coin flip, but the displacement operator simultaneously moves the
walker to another circle in phase space. As $\lambda$ in Eq.~(\ref{eq:HD}) is real, the kick is a displacement in~$x$. The
nature of QW on circles in phase space is made clear in Fig.~\ref{fig:blobby}.

In this geometric representation, the coin flip
Hadamard operation is accompanied by a concomitant displacement that shifts the
walker's distribution (the large black dot in Fig.~\ref{fig:blobby}) from one circle of
radius $n_j$ to another circle of radius $n_{j'}$.
To understand the effect of hopping to different circles of phase space, let us consider a coin+walker state initially in the
state~$|0,\alpha\rangle$ with $\alpha=\frac{x+\mathrm{i}p}{\sqrt{2}}$, which corresponds to the coin in the $0$ state and the
walker localized at $(x,p)$ in phase space.

\begin{figure}[ht]
\includegraphics[width=8.5 cm]{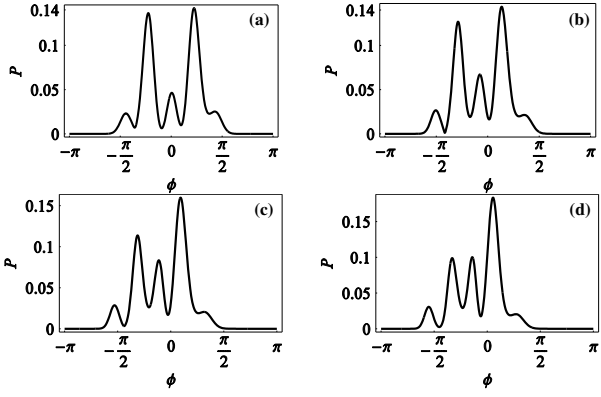}
  \caption{The phase distribution $P(\phi;N,\lambda,\Delta\theta)$ for the walker's location after $N=4$ steps of
   the QW over the different circles in phase space with initial state $(\ket{0,\alpha}+\mathrm{i}\ket{1,\alpha})/\sqrt{2}$,
   $\Delta\theta=0.35$ and (a) $\lambda=0$, (b) $\lambda=0.2$, (c) $\lambda=0.3$ and (d) $\lambda=0.4$.}
   \label{fig:phasedistribution}
\end{figure}

\subsection{The First Step}

The first step corresponds to the application of the unitary operator
\begin{equation}
\label{eq:generalizedU} U=F(H\otimes D).
\end{equation}
First the generalized Hadamard transformation $H\otimes D(\mathrm{i}\lambda/\sqrt{2})$ is applied:
\begin{equation}
\label{eq:HDon0alpha}
    H\otimes D\left|0,\frac{x+\mathrm{i}p}{\sqrt{2}}\right\rangle
        = \left|+,\frac{x+\lambda+\mathrm{i}p}{\sqrt{2}}\right\rangle.
\end{equation}
This generalized Hadamard operator is then followed by the unitary conditional phase operator $F$ on the
state~(\ref{eq:HDon0alpha}), which yields the resultant state
\begin{align}
\label{eq:Uon0alpha}
    U\left|0,\frac{x+\mathrm{i}p}{\sqrt{2}}\right\rangle
        =& \frac{1}{\sqrt{2}}\Bigg[
            \left|0,\frac{x+\lambda+\mathrm{i}p}{\sqrt{2}}\mathrm{e}^{\mathrm{i}\Delta\theta}\right\rangle
                \nonumber   \\  &
            +\left|1,\frac{x+\lambda+\mathrm{i}p}{\sqrt{2}}\mathrm{e}^{-\mathrm{i}\Delta\theta}\right\rangle
            \Bigg].
\end{align}
Eq.~(\ref{eq:Uon0alpha}) has three important features. One is that the resultant state is an entanglement of a coherent state
with a qubit of the type that is observed in microwave cavity quantum electrodynamics experiments~\cite{Har01}. The second
important point is that each of the two walker states
$\left|(x+\lambda+\mathrm{i}p)/\sqrt{2}\mathrm{e}^{\pm\mathrm{i}\Delta\theta}\right\rangle$ are localized on the same circle in phase
space, and third the rotation of the coherent state by angle~$\Delta\theta$ is \emph{independent} of which circle the walker
is on.

Thus, although the walker is forced to hop between circles during the application of each Hadamard
transformation~(\ref{eq:HD}), we will show that the QW survives this generalized action.

\subsection{After N Steps}

Consider an initial state of the coin+walker as
\begin{equation}
\label{eq:initial}
\ket{\Phi(\alpha)}=\frac{1}{\sqrt{2}}\ket{0,\alpha}+\frac{\mathrm{i}}{\sqrt{2}}\ket{1,\alpha})
\end{equation}
The state after N steps is $\ket{\Phi(N)}=U^N\ket{\Phi}$. The phase distribution for the walk after N steps is $P(\phi)$ in
Eq.~(\ref{eq:phasestate}) for $\rho_\text{w}=\text{Tr}_c(\ket{\Phi(N)}\bra{\Phi(N)})$. The first 3 steps for the walk and corresponding phase
distributions are discussed in the Appendix.

\begin{widetext}
The state after N steps is
\begin{equation}
\label{eq:phiN}
\ket{\Phi(N,\alpha,\lambda,\Delta\theta)}=\sum^{2^{N-1}}_{i=1}\big[p_i(N)\ket{0,\alpha_i(N,\alpha,\lambda,\Delta\theta)}
+q_i(N)\ket{1,\beta_i(N,\alpha,\lambda,\Delta\theta)}\big].
\end{equation}
The coin+walker state (\ref{eq:phiN}) adopts a simple form: it is an entanglement between orthogonal coin qubit states with superpositions of
coherent states. The weights ${p_i,p_j}$ and coherent state amplitudes ${\alpha_i,\beta_j}$ are determined by recursion
relations presented in the Appendix. After tracing out the coin state,
\begin{align}
\rho_\text{w}(N,\alpha,\lambda,\Delta\theta)= \sum_{i,j} \Big[p_i(N)
p^*_j(N)\ket{\alpha_i(N,\alpha,\lambda,\Delta\theta)}\bra{\alpha_j(N,\alpha,\lambda,\Delta\theta)}
+q_i(N)q^*_j(N)\ket{\beta_i(N,\alpha,\lambda,\Delta\theta)}\bra{\beta_j(N,\alpha,\lambda,\Delta\theta)}\Big]
\end{align}
is obtained. The phase distribution for the state after N steps is thus
\begin{align}
\label{eq:phase} P(\phi;N,\alpha,\lambda,\Delta\theta)= \lim_{M\rightarrow\infty}\sum_{i,j}&\Big[p_i(N) p^*_j(N)
_M\!\langle\phi|\alpha_i(N,\alpha,\lambda,\Delta\theta)\rangle
\times\langle\alpha_j(N,,\alpha,\lambda,\Delta\theta)|\phi\rangle_M\nonumber\\
&+q_i(N)q^*_j(N)_M\!\langle\phi|\beta_i(N,\alpha,\lambda,\Delta\theta)\rangle
\times\langle\beta_j(N,\alpha,\lambda,\Delta\theta)|\phi\rangle_M\Big]
\end{align}
where the overlap of the phase state with the coherent state given by
\begin{equation}
_M\!\langle\phi|\alpha\rangle= \mathrm{e}^{-|\alpha|^2/2} \frac{1}{\sqrt{M}}\sum_{n=0}^{M-1}\frac{\left(\alpha
\mathrm{e}^{-\mathrm{i}\phi}\right)^n}{\sqrt{n!}},
\end{equation}
which is a function of both $\lambda$ and $\Delta\theta$.
\end{widetext}

\subsection{The Spread in Phase}

In order to observe a QW, the choice of parameters is critical. Therefore, we study how choices of $\Delta\theta$ and
$\lambda$ can affect the quality of the phase distribution for revealing a signature of a QW. We expect that the choice of
$\Delta\theta$ controls the rate of spreading of the phase distribution because $\Delta\theta$ corresponds to the size of the
walker's step. On the other hand, $\lambda$ is responsible for breaking the symmetry of $P(\phi)$ around $\phi=0$. We can see
these effects in Fig.~\ref{fig:phasedistribution}. Specifically, we observe that for increasing $\lambda$, the overall
distribution becomes more skewed towards positive $\phi$ and individual peaks can become narrower. The skewing is due to the
increasing contribution from $\ket{\beta_i}$, which can be higher in amplitude than the $\ket{\alpha_i}$ terms, hence the
concomitant narrowing of some peaks.

\begin{figure}[ht]
\includegraphics[width=8.5 cm]{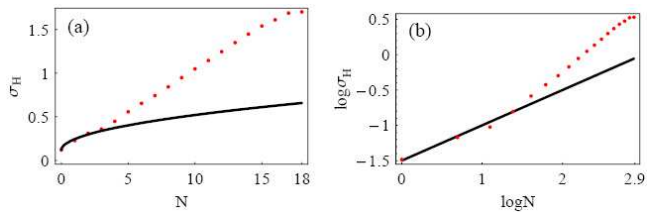}
   \caption{The Holevo standard deviation~$\sigma_\text{H}$ of the phase
distribution for the random and quantum walks, for $\alpha=3$, $ \lambda=0.4$, and $\Delta\theta=0.35$, as a function of the
number of steps N presented as (a)$\sigma_\text{H}$ vs~$N$ and as (b) $\log \sigma_\text{H}$ vs~$\log N$ for the classical
random walk (solid line) and the QW (dots).}
\label{fig:deviation}
\end{figure}

The spread of the phase distribution provides an important signature of the QW, and we use the Holevo standard
deviation~$\sigma_\text{H}$ (\ref{eq:Holevo}) to quantify this spread. The graphs of $\sigma_\text{H}$ vs~$\log N$ and its
log-log version in Fig.~\ref{fig:deviation} clearly reveal the square root spreading feature for the random walk and the
quadratic enhancement for the QW. Therefore, the QW behavior is clearly present despite having generalized the Hadamard
transformation to Eq.~(\ref{eq:HD}) and used a Holevo standard deviation for phase as a quantifier. Fig.~\ref{fig:deviation}
thus makes it clear that the QWs over different circles in phase space are actual QWs.

\subsection{The Photon Number Distribution}

A complication of random and quantum walks over different circles is that the number distribution can vary as the walker is
effectively moving nearer and farther from the origin in phase space with the application of each generalized Hadamard
transformation~(\ref{eq:HD}). This hopping is responsible for the narrowing of individual peaks in
Fig.~\ref{fig:phasedistribution} as discussed earlier.
\begin{widetext}For
\begin{align}
\label{eq:Pn}
    P(n;N,\alpha,\lambda,\Delta\theta)&=
    \Big|\Big\langle n\left|\rho_\text{w} \left(N,\alpha,\lambda,\Delta\theta\right)\ket{n} \right| \nonumber\\
    &=\sum_{i,j}p_i(N)p^*_j(N)\!\langle n|\alpha_i(N,\alpha,\lambda,\Delta\theta)\rangle
    \!\langle\alpha_j(N,\alpha,\lambda,\Delta\theta)|n\rangle+q_i(N)q^*_j(N)\!\langle n
    |\beta_i(N,\alpha,\lambda,\Delta\theta)\rangle\!\langle\beta_j
    (N,\alpha,\lambda,\Delta\theta)|n\rangle
\end{align}
the number distribution, the walker's effective distance from the origin in phase space is given by
\begin{equation}
\label{eq:sqrtbarn}
    \sqrt{\bar{n}}=\sqrt{\sum_{n=0}^\infty nP(n)},
\end{equation}
where $N$, $\alpha$, $\lambda$, and $\Delta\theta$ are suppressed from the expression for brevity, and the walker's radial
spread in phase space is given by
\begin{equation}
\label{eq:deltan}
    \delta n =\sqrt{\left\langle\hat{n}-\bar{n}\right\rangle^2}.
\end{equation}
The expression for $\bar{n}$ is
\begin{align}
\label{eq:photonnumber}
    \bar{n}(N,\alpha,\lambda,\Delta\theta)
    =\sum_{i,j}p_i(N)p^*_j(N)\alpha^*_i(N,\alpha,\lambda,\Delta\theta)\alpha_j(N,\alpha,\lambda,\Delta\theta)
    +q_i(N)q^*_j(N)\beta^*_i(N,\alpha,\lambda,\Delta\theta)\beta_j(N,\alpha,\lambda,\Delta\theta),
\end{align}
which can be approximated by
\begin{align}
\label{eq:approxeqphotonnumber} \bar{n}(N,\alpha,\lambda,\Delta\theta)\approx
-\frac{1}{2}\Big\{-(\alpha+\lambda)^2+\alpha(\alpha+\lambda)\cos{\Delta\theta} +\lambda(\alpha+\lambda)\cos{N\Delta\theta}
-\frac{\alpha\lambda\cos{[\Delta\theta(1-N)]}}{\sin^2{(\Delta\theta/2)}}\Big\}.
\end{align}
for large~$N$.
\end{widetext}
Eqs.~(\ref{eq:photonnumber}) and~(\ref{eq:approxeqphotonnumber}) for the mean number and (\ref{eq:deltan}) for the spread of
the walker quantify the degree of hopping between circles in phase space, and these expressions will be useful in the next
section. Although there is hopping to different circles, the QW is clearly evident in the quadratic enhancement of phase
spreading, with respect to the Holevo standard deviation, shown in Fig.~\ref{fig:deviation}. Thus, provided that the
parameters~$\alpha$, $\lambda$, and~$\Delta\theta$ are chosen judiciously, the generalization of the Hadamard coin flip
transformation from (\ref{eq:H}) to (\ref{eq:HD}) does not destroy the QW, but it does modify the QW from being on a
\emph{circle} in phase space to being on \emph{circles} in phase space. In the next section, we approach the generalized
Hadamard transformation from the microscopic perspective, and the mean number~$\bar{n}$ turns out to be important with respect
to controlling the QW in order to ensure optimal enhancement of phase spreading.

\section{From Jaynes-Cummings evolution to quantum walks}
\label{sec:JC2QW}

In the previous section, we treated the indirectly driven coin via the generalized Hadamard transformation~(\ref{eq:HD}), but
this transformation was introduced by fiat. In this section we consider the JC model Hamiltonian \cite{JC}, which underpins so
much of quantum optics and cavity quantum electrodynamics, as a foundation for obtaining the generalized Hadamard
transformation, or at least a good approximation to this transformation under reasonable conditions.

In quantum optics, the simple harmonic oscillator is typically the single mode electromagnetic field within the cavity, and
the coin is an atom transiting the cavity. Cavity quantum electrodynamic realizations of QWs on the circle in phase space have
been suggested~\cite{San03,Di04}.

\subsection{Driven Jaynes-Cummings Model With Large Detuning}

For a simple harmonic oscillator with angular resonant frequency~$\omega_\text{r}$, coupled with strength~$g$ to a qubit of
angular resonant frequency $\omega_\text{a}$, the JC dynamics for the joint system is given by~\cite{JC}
\begin{equation}
\label{eq:JCHamiltonian}
    \hat{H}_\text{JC}=\omega_\text{r} \left(\hat{n}+1/2 \right)
        +\frac{\omega_\text{a}}{2}
        \hat{\sigma}_z+g(\hat{a}^\dagger\hat{\sigma}_{-}+\hat{a}\hat{\sigma}_{+}).
\end{equation}
The joint system is driven by a time-dependent driving force (or field) by directly
driving the simple harmonic oscillator according to
\begin{equation}
    \hat{H}_\text{dr}=\epsilon(t)\left(\hat{a}^\dagger \mathrm{e}^{-\mathrm{i}\omega_\text{d}t}
        + \hat{a}\mathrm{e}^{\mathrm{i}\omega_\text{d}t}\right)
\end{equation}
with $\epsilon(t)$ the amplitude and $\omega_\text{d}$ the driving carrier frequency. For simplicity we let $\epsilon(t)$ be a
constant $\epsilon$ for some of the time and zero for other times.

For large detuning $g\ll|\Delta|=|\omega_\text{a}-\omega_\text{r}|$, conjugating the JC Hamiltonian under the action of
\begin{equation}
\label{eq:V}
    V=\exp\left[\frac{g}{\Delta}(\hat{a}^\dagger\hat{\sigma}_--\hat{a}\hat{\sigma}_+)\right]
\end{equation}
yields the effective Hamiltonian
\begin{align}
    \tilde{\hat{H}}_\text{JC}=&V\hat{H}_\text{JC}V^\dagger \nonumber \\
    \approx & (\omega_\text{r}+\chi\hat{\sigma}_z)\hat{n}
    +\frac{1}{2}(\omega_\text{a}+\chi)\hat{\sigma}_z
    +O(\chi^2)
\end{align}
for $\chi\equiv g^2/\Delta$; the conjugated driving Hamiltonian is thus
\begin{equation}
\label{eq:dr}
\tilde{\hat{H}}_\text{dr}
    = V\hat{H}_\text{dr}V^\dagger 
    \approx\epsilon(t)\left[\left(\hat{a}+\frac{g}{\Delta}\hat{\sigma}_-\right)^\dagger
        \mathrm{e}^{-\mathrm{i}\omega_\text{d}t}+\text{hc}\right]
\end{equation}
for `hc' designating the Hermitian conjugate. The time evolution of Eq.~(\ref{eq:dr}) leads to the generalized Hadamard
transformation (\ref{eq:HD}).

\subsection{Implementation of The Generalized Hadamard Transformation}

To implement a QW, first we turn on the driving force $(\epsilon(t)=\epsilon)$ for the Hadamard transformation. In a frame
rotating at the drive frequency $\omega_\text{d}$, the effective Hamiltonian of the coin+walker system is thus
\begin{align}
    \hat{H}_\text{eff} =&\frac{1}{2}\left[2\chi\left(\hat{n}+1/2\right)-\delta_\text{da}\right]\hat{\sigma}_z
  -\delta_\text{dr}\hat{n}\nonumber\\
&+\frac{\Omega_\text{R}}{2}\hat{\sigma}_{x}+\epsilon(\hat{a}^\dagger + \hat{a}),
\label{eq:Heff}
\end{align}
with the detuning of the qubit transition frequency from the driving
force
\begin{equation}
\label{eq:omegada} \delta_\text{da}=\omega_\text{d}-\omega_\text{a},
\end{equation}
the detuning of the resonator from the
driving force
\begin{equation}
\label{eq:deltadr} \delta_\text{dr}=\omega_\text{d}-\omega_\text{r},
\end{equation}
and the Rabi frequency
\begin{equation}
\label{eq:rabi}
\Omega_\text{R}=2g\epsilon/\delta_\text{dr}.
\end{equation}
The first term in Eq.~(\ref{eq:Heff}) expression effects the coin-induced walker phase shift. The unitary operator generated
by the effective Hamiltonian $\hat{H}_\text{eff}$ is
\begin{equation}
\label{eq:unitaryH}
\exp\left[-\mathrm{i}\hat{H}_\text{eff}t_\text{H}\right]=(H\otimes D)\Xi,
\end{equation}
which is a good approximation to the generalized Hadamard transformation in (\ref{eq:HD}) for $D(\alpha=-\mathrm{i}\epsilon
t_\text{H})$ the displacement operator~(\ref{eq:displacement}) and $\Xi$ is a `small' operator explicitly shown in
Eq.~(\ref{eq:Xi0}).

Choosing \cite{Bla04} $\omega_\text{d}=2\bar{n} \chi+\omega_\text{a}$, $\hat{H}_\text{eff}$ then generates rotations of the
qubit about the $x$ axis with Rabi frequency $\Omega_\text{R}$. In particular, choosing
\begin{equation}
\label{eq:omegad} \omega_\text{d}=2\bar{n}\chi-2g\epsilon/\Delta+\omega_\text{a}
\end{equation} and
\begin{align}
\label{eq:th}
t_\text{H}&=\pi/2\Omega_\text{R}\nonumber\\&=\frac{\pi}{4g\epsilon}\big[\Delta+2\bar{n}\chi-2g\epsilon/\Delta\big]
\end{align}
generates the Hadamard transformation for the coin state
\begin{equation}
H=\mathrm{e}^{\mathrm{i}t_\text{H}\Omega_\text{R}/2\hat{\sigma}_x}
\end{equation} within the generalized Hadamard transformation (\ref{eq:unitaryH}). The choice of pulse duration~$t_\text{H}$ is
critical in effecting a Hadamard transformation, but this duration itself is a function of~$\bar{n}$, which we know from the
previous section is time-dependent because the walker is hopping between circles in phase space. Specifically $t_\text{H}$
depends inversely on~$\Omega_\text{R}$~(\ref{eq:rabi}), which is itself inversely proportional to the driving field detuning
$\delta_\text{dr}$ (\ref{eq:deltadr}). The driving field detuning is a function of~$\omega_\text{d}$ (\ref{eq:omegad}),
and~$\omega_\text{d}$ is dependent on~$\bar{n}$ (\ref{eq:omegad}). Therefore, the duration of each pulse~ $t_\text{H}$ must be
chosen in accordance with the value of~$\bar{n}$ for the system.

In order to choose the appropriate pulse duration~$t_\text{H}$ for each step, we employ the following protocol, which depends
on the time-dependent mean number~$\bar{n}$. In this protocol, $\bar{n}$ is obtained from a theoretical analysis rather than
continuous measurements or sampling, which could disturb the system. In the first step we let
\begin{equation}
\label{eq:firststep-barn}
    \bar{n}=|\alpha|^2
\end{equation}
and use this value to determine~$\omega_\text{d}$ according to Eq.~(\ref{eq:omegad}). Then this value of~$\omega_\text{d}$ is
used to compute $ \Omega_\text{R}$ and, from this, $t_\text{H}$. The duration of the generalized Hadamard pulse is precisely
this value of~$t_\text{H}$. In subsequent steps $\bar{n}$ will have changed due to the walker hopping to other circles in
phase space, so $\bar{n}$ has to be computed and used in a protocol described in Subsection~IV.D.

We now have expressions for $H$ and $D$ in Eq.~(\ref{eq:unitaryH}) and require
\begin{widetext}
\begin{align}
\label{eq:Xi0}
    \Xi=\prod_{n=0}^\infty &\exp\Bigg[\frac{-t_\text{H}}{2}(-\mathrm{i}
        t_\text{H}\chi\hat{n})^{2n+1}\Omega_\text{R}\hat{\sigma}_y
        +\frac{\mathrm{i}t_\text{H}}{2}(-\mathrm{i}t_\text{H}\chi)^{2n+1}\epsilon(\hat{a}^\dagger-\hat{a})\hat{\sigma}_z\Bigg]
            \nonumber   \\  &
        \times\exp\Bigg[\frac{\mathrm{i}t_\text{H}}{2}(-\mathrm{i}t_\text{H}\chi\hat{n})^{2n+2}
        \Omega_\text{R}\hat{\sigma}_x
        +\frac{\mathrm{i}t_\text{H}}{2}(-\mathrm{i}t_\text{H}\chi)^{2n+2}\epsilon(\hat{a}^\dagger+\hat{a})\Bigg].
\end{align}
\end{widetext}

We can see that $\Xi$ is close to unity for our choice of parameters; thus Eq.~(\ref{eq:unitaryH}) tends to the generalized
Hadamard of Eq.~(\ref{eq:HD}). The spectrum of the operators $\hat{\sigma}_z$ and $\hat{\sigma}_x$ is ${0,1}$, and the
relative size of $\hat{a}^\dagger+\hat{a}$ and $\mathrm{i}(\hat{a}^\dagger-\hat{a})$ is never much more than $|\alpha|$
because $|\langle \hat{\alpha}\rangle|=|\alpha|$. In the case $t_\text{H}\chi=\pi g/4\epsilon\ll 1 $, we neglect the higher
orders of $t_\text{H}\chi$. In the case of large detuning, that is $g/\delta_\text{dr}\ll 1$, the term
\begin{equation}
-\frac{t_\text{H}}{2}(-\mathrm{i}t_\text{H}\chi\hat{n})\Omega_\text{R}\hat{\sigma}_y=\mathrm{i}\frac{t_\text{H}\pi
g^2}{4\delta_\text{dr}}\hat{n}\hat{\sigma}_y
\end{equation}
can also be neglected. Thus $\Xi$ can be approximated by
\begin{equation}
\label{eq:Xi}
    \Xi\approx\exp\left[\frac{\pi}{8}t_\text{H}g(\hat{a}^\dagger-\hat{a})\hat{\sigma}_z\right].
\end{equation}
The evolution of initial states under $\Xi$ are shown as
\begin{equation}
\label{eq:Xionstate}
    \Xi\ket{j,\alpha}\approx \ket{j,\alpha+(-1)^j \pi t_\text{H}g/8},j=0,1.
\end{equation}
As~$\Xi$ in Eq.~(\ref{eq:Xi}) is close to an identity operation for the restricted choices of parameters, the resultant
generalized Hadamard transformation~(\ref{eq:unitaryH}) is quite close to the ideal~(\ref{eq:HD}) in the previous section. It
is thus important to choose parameters for which~ $\Xi$ can be neglected. In this case, the displacement operator~$D$ in
Eq.~(\ref{eq:unitaryH}) is responsible for displacing the walker's distance from the origin in phase space by $|\alpha|
\mapsto |\alpha|(1+\epsilon t_\text{H}/2)$. Fortunately, even the effects of this induced jump in~$|\alpha|$ can be minimized
by varying the duration of successive generalized Hadamard pulses.

\subsection{Implementation of The First Step}

In the previous subsection, we have seen how the generalized Hadamard transformation generated by the JC model is very close
to the ideal Hadamard transformation of Sec.~III. The importance of choosing the appropriate duration of the generalized
Hadamard pulse was noted in Subsection~IV.B. Each step of the QW corresponds to first performing the generalized Hadamard
transformation and then the conditional phase shift operation given by~$F$ (\ref{eq:F}). In this subsection we concentrate
solely on the walker's first step, which is the generalized Hadamard transformation followed by~$F$.

The conditional phase shift~$\Delta\theta$ has a size that is constrained by (\ref{eq:deltatheta}). In terms of parameters in
the JC model, the step size is
\begin{equation}
\label{eq:DeltathetagtautH}
    \Delta\theta=\pm\chi(\tau+t_\text{H}),
\end{equation}
for $\tau$ the time between generalized Hadamard pulses. Because the JC Hamiltonian applies to the dynamics both during the
generalized Hadamard pulse, which has duration~$t_\text{H}$, and during the period between these pulses, which has
duration~$\tau$, the step size~ (\ref{eq:DeltathetagtautH}) is proportional to the total time for each step,
namely~$\tau+t_\text{H}$.

At time $\tau+t_\text{H}$ the first step is completed, but~$\bar{n}$ has changed. The new~$\bar{n}$ after the completion of
the first step is required to calculate the appropriate $t_\text{H}$ for the second step. The value of $\bar{n}$ after the
first step is readily obtained from Eq.~(\ref{eq:photonnumber}) by inserting the relevant parameters as well as $N=1$. From
this value of~$\bar{n}$, the pulse duration for the next generalized Hadamard transformation is given by
Eq.~(\ref{eq:photonnumber}). This knowledge of $t_\text{H}$ for the next generalized Hadamard transformation prepares us for
the second step.


\subsection{N Steps}

The previous subsection describes how to perform the first step and obtain the information required to set the duration for
the subsequent generalized Hadamard transformation. In this subsection we describe the transformations required for the walker
to go an arbitrary number~$N$ steps. Unlike the case of the quantum walk on a single circle or the case of quantum walks on
circles described in Sec.~III, here the choice of $t_\text{H}$ for each circle is more complicated but quite important.

\begin{figure}[ht]
\includegraphics[width=8.5 cm]{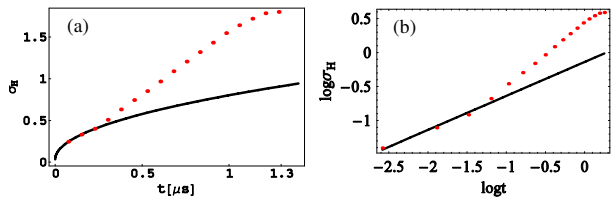}
   \caption{
   (a) The Holevo Standard deviation of phase for both the quantum and classical random walks up to $N=15$
   for $\alpha=3$, $d=21$, and typical system parameters
   $(\omega_\text{a},\omega_\text{r}, g,\epsilon)/2\pi=(7000, 5000, 100, 1000)$ MHz~\cite{Wal05}.
   Numerical simulations reveal that the Holevo standard deviation is almost independent of the initial state of
   charge qubit, and is approximately linear in $N$, $\sigma_\text{H}=(1.3964\pm0.0180)t+(0.1208\pm0.0018)$.
   (b) The numerically simulated Holevo standard deviation for phase distribution in log-log scale
   is shown to be approximately linear in $\log{t}$: $\log{\sigma_\text{H}}=(0.924\pm0.009)\log{t}+(0.442\pm0.004)$ and
   the $r$ coefficient is 0.99.}
\label{fig:numerical}
\end{figure}

For an arbitrary $i^\text{th}$ step, we can calculate the average photon number $\bar{n}(i,\alpha,\lambda,\Delta\theta)$ based
on the analytical result (\ref{eq:photonnumber}), and then decide the pulse duration of the $i^\text{th}$ step (\ref{eq:th})
\begin{equation}
    t^i_\text{H}=\frac{\pi}{4g\epsilon}
        \left[\Delta+2\bar{n}(i,\alpha,\lambda,\Delta\theta)\chi-2g\epsilon/\Delta\right].
\end{equation}
We apply the generalized Hadamard transformation $\exp{\left[-\mathrm{i}\hat{H}_\text{eff}t^i_\text{H}\right]}$ following by
the unitary operator of the free evolution $\exp{\big[\mathrm{i}\chi(\tau+t^i_\text{H})\big]}\hat{n}\hat{\sigma}_z$. These two
applications together effect unitary operation
\begin{equation}
    U_\text{eff}\approx F(H\otimes D).
\end{equation}

Using our protocol for choosing durations of generalized Hadamard pulses, we obtain numerically the Holevo standard deviation
for the phase distribution of the reduced walker state as a function of time~ $t$. In contrast to the related plots in
Fig.~\ref{fig:deviation} of Sec.~III, which depend on the number of steps~$N$, these plots explicitly depend on~$t $. In
Sec.~III the choice of $N$ vs $t$ is not significant because $t \propto N$; here, however, $t$ is not proportional to~$N$
because of the varying duration of each step due to the variability of $\bar{n} $. In physical systems, the random walk is
characterized by its time dependence so, in that spirit, we also use time~$t$, rather than the number of pulses $N$, to show
the quadratic enhancement of the phase spreading for the QW vs the random.

This quadratic enhancement is evident in Fig.~\ref{fig:numerical}. To show this more explicitly we apply linear regression
techniques to the log-log plot, which theoretically should be linear with a slope of $1/2$ in the classical case (depicted as
a solid line in Fig.~\ref{fig:numerical}(b)) and slope~$1$ in the quantum case for small spreading of phase. The linear
regression results are presented in detail in the caption of Fig.~\ref{fig:numerical}, and residual $r=0.99$ for the QW,
confirming the linear relationship between $\sigma_\text{H}$ vs~$t$. The slope is $0.924$, which is quite close to unity.
Together the slope being close to unity and the high value of~$r$ demonstrate that this protocol does indeed lead to an
enhancement of phase spreading that is very close to quadratic and is thus a signature of QW behavior.

\section{Conclusions}
\label{sec:conclusions}

Motivated by the physical difficulty of directly driving a coin qubit in a cavity quantum electrodynamical realization of the
quantum quincunx, we generalized the Hadamard coin flip to also kick the resonator. In Sec.~III this kick was incorporated
within an idealized generalized Hadamard transformation, and Sec.~IV approximately obtained the generalized Hadamard
transformation directly from the ubiquitous Jaynes-Cummings Model.

The generalization of the Hadamard transformation modifies the walk from being on one circle in phase space to hopping between
circles in phase space. Despite this hopping, the quantum walk is evident, in the quadratically enhanced spreading of phase.
In Sec.~IV the duration of each generalized Hadamard pulse is modified according to which circle the walker is
on---equivalently the time-dependent mean number~$\bar{n}$---which means that the spreading of phase in time~$t$ is slightly
different from spreading as a function of number of steps~$N$. We show the quadratic enhancement in terms of the more
experimentally relevant time~$t$, which is the signature of quantum walk behavior.

As explained in \cite{San03}, the quantum walk behavior can be ascertained by bringing in controllable decoherence. Then
tuning of decoherence will interpolated the phase spreading from linear in time to the square root of time. Furthermore,
although phase is not directly measured, its cosine and sine can be inferred from homodyne measurements, or from full optical
homodyne tomography.

\begin{appendix}
\begin{center}\bf{Appendix}\end{center}
We calculate the state of the coin+walker system after N steps for N small. For the initial state $\ket{\Phi}$ in
Eq.~(\ref{eq:initial}), after the $N^\text{th}$ step of walking on the circles, the state $\ket{\Phi(N)}=U^N\ket{\Phi}$ is
shown in Eq.~(\ref{eq:phiN}), where the coefficients $p_i(N)$ and $q_i(N)$ are obtained from the following recursion relations
(for $N\ge 2$)
\begin{equation}
p_i(N)=
\begin{cases}
p_i(N-1)/\sqrt{2} & \text{if } 1\le i\le 2^{N-2} \\
q_{i-2^{N-2}}(N-1)/\sqrt{2} & \text{if } 2^{N-2}<i\le 2^{N-1},\tag{A1}
\end{cases}
\end{equation}
and\begin{equation} q_i(N)=
\begin{cases}
p_i(N-1)/\sqrt{2} & \text{if } 1\le i\le 2^{N-2} \\
-q_{i-2^{N-2}}(N-1)/\sqrt{2} & \text{if } 2^{N-2}<i\le 2^{N-1},\tag{A2}
\end{cases}
\end{equation}
For the case $N=0$, we have $p_0(0)=1/\sqrt{2}$ and $q_0(0)=-\mathrm{i}/\sqrt{2}$. We will show the case $N=1$ below.

The coherent state with $\alpha_i(N,\lambda,\Delta\theta)$ and $\beta_i(N,\lambda,\Delta\theta)$ can also obtained from the
following recursion relations for $N\ge 1$
\begin{widetext}
\begin{equation}
\alpha_i(N,\lambda,\Delta\theta)=
\begin{cases}
\left[\alpha_i(N-1,\lambda,\Delta\theta)+\lambda\right]e^{\mathrm{i}\Delta\theta} & \text{if } 1\le i\le 2^{N-2} \\
\left[\beta_{i-2^{N-2}}(N-1,\lambda,\Delta\theta)+\lambda\right]e^{\mathrm{i}\Delta\theta} & \text{if } 2^{N-2}<i\le
2^{N-1},\tag{A3}
\end{cases}
\end{equation}
and\begin{equation} \beta_i(N,\lambda,\Delta\theta)=
\begin{cases}
\left[\alpha_i(N-1,\lambda,\Delta\theta)+\lambda\right]e^{-\mathrm{i}\Delta\theta} & \text{if } 1\le i\le 2^{N-2} \\
\left[\beta_{i-2^{N-2}}(N-1,\lambda,\Delta\theta)+\lambda\right]e^{-\mathrm{i}\Delta\theta} & \text{if } 2^{N-2}<i\le
2^{N-1},\tag{A4}
\end{cases}
\end{equation}
\end{widetext}
For the case $N=0$, $\alpha_0(0)=\beta_0(0)=\alpha$.

After the first step, the state of the system is
\begin{align}
\ket{\Phi(1)}=p_1(1)\ket{0,\alpha_1(1)}+q_1(1)\ket{1,\beta_1(1)},\tag{A5}
\end{align}
with
\begin{align}
&p_1(1)=\frac{1+\mathrm{i}}{2}, q_1(1)=\frac{1-\mathrm{i}}{2}, \tag{A6}\\
&\alpha_1(1)=(\alpha+\lambda)e^{\mathrm{i}\Delta\theta},\beta_1(1)=(\alpha+\lambda)e^{-\mathrm{i}\Delta\theta}. \nonumber
\end{align}
After the second step, the state is
\begin{align}
\ket{\Phi(2)}=\sum_{i=1}^2p_i(2)\ket{0,\alpha_i(2)}+q_i(2)\ket{1,\beta_i(2)}, \tag{A7}
\end{align}
with
\begin{align}
&p_1(2)=\frac{1+\mathrm{i}}{2\sqrt{2}}, p_2(2)=\frac{1-\mathrm{i}}{2\sqrt{2}},\tag{A8} \\
&q_1(2)=\frac{1+\mathrm{i}}{2\sqrt{2}}, q_2(2)=-\frac{1-\mathrm{i}}{2\sqrt{2}},\nonumber\\
&\alpha_1(2)=\alpha e^{2\mathrm{i}\Delta\theta}+\lambda(e^{2\mathrm{i}\Delta\theta}+e^{\mathrm{i}\Delta\theta}), \nonumber\\
&\alpha_2(2)=\alpha+\lambda(e^{\mathrm{i}\Delta\theta}+1),\nonumber \\
&\beta_1(2)=\alpha+\lambda(1+e^{-\mathrm{i}\Delta\theta}),\nonumber\\
&\beta_2(2)=\alpha e^{-2\mathrm{i}\Delta\theta}+\lambda(e^{-\mathrm{i}\Delta\theta}+e^{-2\mathrm{i}\Delta\theta}).\nonumber
\end{align}
The third step leads the state to
\begin{align}
\ket{\Phi(3)} =\sum_{i=1}^4p_i(3)\ket{0,\alpha_i(3)}+q_i(3)\ket{1,\beta_i(3)}, \tag{A9}
\end{align}
with
\begin{align}
&p_1(3)=\frac{1+\mathrm{i}}{4}, p_2(3)=\frac{1-\mathrm{i}}{4}, \tag{A10}\\
&p_3(3)=\frac{1+\mathrm{i}}{4}, p_4(3)=-\frac{1-\mathrm{i}}{4}, \nonumber \\
&q_1(3)=\frac{1+\mathrm{i}}{4}, q_2(3)=\frac{1-\mathrm{i}}{4},  \nonumber \\
&q_3(3)=-\frac{1+\mathrm{i}}{4}, q_4(3)=\frac{1-\mathrm{i}}{4}, \nonumber \\
&\alpha_1(3)=\alpha e^{3\mathrm{i}\Delta\theta}+\lambda(e^{3\mathrm{i}\Delta\theta}+e^{2\mathrm{i}\Delta\theta}+e^{\mathrm{i}\Delta\theta}), \nonumber\\
&\alpha_2(3)=\alpha e^{\mathrm{i}\Delta\theta}+\lambda(e^{2\mathrm{i}\Delta\theta}+2e^{\mathrm{i}\Delta\theta}),\nonumber \\
&\alpha_3(3)=\alpha e^{\mathrm{i}\Delta\theta}+\lambda(2e^{\mathrm{i}\Delta\theta}+1),\nonumber\\
&\alpha_4(3)=\alpha e^{-\mathrm{i}\Delta\theta}+\lambda(e^{\mathrm{i}\Delta\theta}+1+e^{-\mathrm{i}\Delta\theta}),\nonumber \\
&\beta_1(3)=\alpha e^{\mathrm{i}\Delta\theta}+\lambda(2e^{\mathrm{i}\Delta\theta}+e^{-\mathrm{i}\Delta\theta}), \nonumber\\
&\beta_2(3)=\alpha e^{-\mathrm{i}\Delta\theta}+\lambda(1+2e^{-\mathrm{i}\Delta\theta}), \nonumber \\
&\beta_3(3)=\alpha e^{-\mathrm{i}\Delta\theta}+\lambda(2e^{-\mathrm{i}\Delta\theta}+e^{-2\mathrm{i}\Delta\theta}), \nonumber\\
&\beta_4(3)=\alpha
e^{-3\mathrm{i}\Delta\theta}+\lambda(e^{-\mathrm{i}\Delta\theta}+e^{-2\mathrm{i}\Delta\theta}+e^{-3\mathrm{i}\Delta\theta}).
\nonumber
\end{align}

\begin{figure}[ht]
\includegraphics[width=8.5 cm]{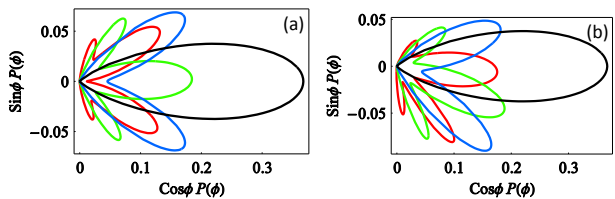} 
   \caption{The phase distribution for the walkers' location after the first three steps of the QW on the different circles with
   initial state $(\ket{0}+\mathrm{i}\ket{1})\ket{\alpha=3}/\sqrt{2}$, $\Delta\theta=0.35$ and different $\lambda$. (a) $\lambda=0$ and (b) $\lambda=0.4$. The yellow line is for the case N=0, the blue one for N=1, the green one for N=2 and the red one for N=3.}
\label{fig:appendix}
\end{figure}

The entanglement between the coin qubit and the superposition of coherent states leads to the signature of QW compared to
random walk, that is the quadratic in phase spreading. From Fig.~\ref{fig:appendix}, for the given $\alpha$ and fixed
$\Delta\theta$, the hopping between circles, i.e.~$\lambda$ leads the phase distribution to be skewed towards positive $\phi$ and individual peaks can become narrower or broader. However, for the case
$\lambda\ll \alpha$, we still obtain the characteristic quadratic enhancement in phase spreading for QW.
\end{appendix}

\begin{acknowledgments}
We are grateful to A. Blais for numerous helpful comments and suggestions. This work has been supported by NSERC, MITACS,
CIFAR, QuantumWorks and iCORE.
\end{acknowledgments}

\end{document}